\documentstyle[preprint,aps]{revtex}   
\begin{document}
\draft

\title{Renormalization group approach to an Abelian sandpile model 
\\ on planar lattices}

\author{Chai-Yu~Lin$^*$ and Chin-Kun~Hu$^{+}$ }

\address{Institute of Physics, Academia Sinica,
Nankang, Taipei 11529, Taiwan}

\maketitle
\begin{abstract}
One important step in the renormalization group (RG) approach to 
a lattice sandpile model is the exact enumeration of all possible
toppling processes of sandpile dynamics inside a cell for
RG transformations. Here we propose a computer algorithm to 
carry out such exact enumeration for cells of planar lattices
in RG approach to Bak-Tang-Wiesenfeld sandpile model
[Phys. Rev. Lett. {\bf 59}, 381 (1987)]
and consider both the reduced-high RG equations 
proposed by Pietronero, Vespignani, and Zapperi (PVZ) [Phys. Rev. Lett.
{\bf 72}, 1690 (1994)] and the real-height RG equations proposed by
Ivashkevich [Phys. Rev. Lett. {\bf 76}, 3368 (1996)]. Using this 
algorithm we are able to carry out RG transformations more quickly with 
large cell size, e.g. $3 \times 3$ cell for the square (sq) lattice in PVZ RG 
equations, which is the largest cell size at the present, and find
some mistakes in a previous paper [Phys. Rev. E {\bf 51}, 1711 (1995)].
For sq and plane triangular (pt) lattices, we obtain 
the only attractive fixed point for each lattice and calculate the
avalanche exponent $\tau$ and the dynamical exponent $z$.
Our results suggest that the increase of the cell size in the
PVZ RG transformation does not lead to more accurate results.
The implication of such result is discussed.
\end{abstract}

\vskip 5 mm

\pacs{05.40-a, 05.50, 05.60Cd, 89.75Da}

\section{Introduction}

In 1987, Bak, Tang and Wiesenfeld (BTW) \cite{bak87} proposed
the concept of self-organized criticality (SOC) in order to understand
the automatic (i.e. without a tuning parameter, such as temperature)
appearance of abundant self-similar structures and scaling quantities
in nature.  BTW also proposed a lattice sandpile model and used Monte
Carlo simulations to simulate this model on square and simple cubic 
lattices. They did observe self-similar structures and scaling quantities
in the simulation data without tuning any parameter.
Since 1987, many natural phenomena have been related to SOC, such as
earthquakes \cite{earth}, forest fires \cite{ff}, 
biological evolution \cite{evo}, rice pile dynamics \cite{oslo1},
turbulence \cite{ms02}, etc.
Many lattice models have also been proposed to illustrate the
behavior of SOC or avalanche processes \cite{cp99,piph01}.
It has been found that the BTW's sandpile model is Abelian \cite{dhar90} and 
some quantities for this model could be calculated exactly 
\cite{dhar90,priezzhev94,dhar99,hilp00}. The
BTW's Abelian sandpile model (ASM) \cite{bak87} has been considered to be
a prototypical model for SOC. Many ideas about behavior of SOC models,
such as universality and scaling \cite{lh01a}, or new
methods for studying SOC models, such as renormalization group theory
\cite{pietronero94,ves95,ivashkevich96,papoyan97}, are often first tested
 in the BTW's ASM.
In the present paper, we propose a computer algorithm which is useful
for carrying out renormalization group transformations for the BTW's ASM.

The BTW's ASM on a lattice $\Re$ of $N$ sites is defined as follows.
Each site of $\Re$ is assigned a height integer; the $i$-th site is assigned
$z_i$ for $1 \le i \le N$. In the beginning of the simulation, 
the height integer at each site is randomly chosen to be 0, 1,\dots, or $z_c-1$,
 where the critical height is $z_c-1$ and $z_c$ 
is the coordination number of the lattice.
For each time interval $T_a$, one particle falls on a randomly chosen
lattice site, say the $i$-th site; the height $z_i$ then becomes $z_i +1$. 
If the new $z_i < z_c$, then randomly choose a site again, say the $k$-th 
site, to add
a particle; if $z_i \ge z_c$, then
the $i$-th site topples and its height $z_i$ becomes $z_i-z_{c}$.
At the same time, each of the nearest neighbor (nn) sites of the $i$-th site
receives one particle, i.e., $z_{\omega(i,j)} \rightarrow z_{\omega(i,j)}+1$,
$\forall j$ where $\omega(i,j)$ is the label of the
$j$th nn site of the $i$-th site. 
This relaxation procedure takes time $t_w$ and we assume
$t_w/T_a \rightarrow 0$.  
If some of the new $z_{\omega(i,j)}$, $\forall$ $j$, 
are equal or larger than $z_c$ again, 
these sites are denoted by $\omega(i,j')$.
Then, the toppling process continues in parallel for each $j'$ 
with $z_{\omega(i,j')} \rightarrow z_{\omega(i,j')}-z_{c}$ and 
$z_{\omega(\omega(i,j'),k)} \rightarrow z_{\omega(\omega(i,j'),k)}+1 $
where $\omega(\omega(i,j'),k)$ is the $k$-th nn site 
of the $j'$-th nn site of the $i$-th site.   The relaxation time for 
these parallel toppling and receiving processes between
$\omega(i,j')$ and $\omega(\omega(i,j'),k)$ takes another time $t_w$.
Usually, the open boundary conditions are used so that when a 
boundary site topples, the particle can leave the system.
The dynamical process continues until the heights of all sites
are less than $z_c$.
In general, if the last toppling site is 
$\omega(\omega(...(\omega(\omega(i,i_1),i_2),i_3)...),i_n)$, the total
toppling process takes time $n \times t_w$.
In this way, a series of toppling processes with toppling area $s$ 
(i.e., the total number of toppled sites)
and relaxation time $n t_w \equiv t$
appears and forms an avalanche which has no
characteristic size.

After repeating many times the process of adding one particle on a randomly
chosen site with subsequent relaxation when the height of the site is equal or
larger then $z_c$, we can obtain a distribution of toppling area $P(s)$
and calculate the average relaxation time $<t>$ for avalanches with
toppling area $s$. It has been found that $P(s) \sim s^{-\tau}$ with
the avalanche exponent $\tau$ and $<t> \sim s^{z/2}$ 
with the dynamical exponent $z$. Manna \cite{manna90} used Monte Carlo simulations
to calculate $\tau=1.22$ and $z=1.21$ for the BTW model on the square (sq) lattice.
Majumdar and Dhar conjectured that $z = 5/4$ which is consistent
 with their own numerical simulations \cite{majumdar92}, and
Priezzhev et al. \cite{pki96} proposed that $\tau=5/4$. By scaling
argument, Tebaldi et al. \cite{tebaldi99} suggested $\tau=6/5$.
Many investigations have been just focused on numerical simulations 
or exact results for height probabilities and exponents for 
the sq lattice and there is little attention to other kinds
of lattices. It is not clear whether the sandpile model on two 
dimensional lattices have the same set of critical exponents.  

The renormalization group (RG) theory has been used successfully to
calculate critical exponents, order parameters, etc, of ordinary phase
transition models, e.g. the Ising model \cite{hu82}, 
the Potts model \cite{hu92a} 
and the percolation model \cite{hu92b,hu95,lh98};
it has also been used to understand universality and scaling in ordinary
phase transition models.  It is of interest to know whether one can also
use RG theory to calculate critical exponents for lattice SOC models,
e.g. the BTW sandpile model. 
In 1994, Pietronero, Vespignani, and Zapperi (PVZ) \cite{pietronero94}
first proposed a RG theory to calculate critical exponents of the BTW
sandpile model on
the sq lattice. They used a stationary
condition to provide a feedback mechanism that drives the system to its
critical state. In fact, in their approach,
the height of a site $z_i$ is reduced to only three cases corresponding
to three classes: stable for $z_i < z_c-1 $, critical for
$z_i =z_c-1$, and unstable for $z_i \ge z_c$. 
The obtained renormalization group equations allow them to
get an attractive fixed point. At this fixed point, they
obtained avalanche exponent
$\tau$ =1.253 and dynamical exponent $z$ =1.168.
In 1995, PVZ \cite{ves95} described in more details their RG approach to 
the BTW  sandpile model.
In 1996, Ivashkevich \cite{ivashkevich96} generalized PVZ's
RG scheme with real height parameters $z_i$ by kinetic equations and used the 
technique of generating function to construct RG equations to obtain the exponent
 $\tau$ =1.248 and height probabilities. In addition, Ivashkevich found that 
for the sq lattice critical particle transfer probabilities 
are close to branching probabilities for
spanning tree. However, Papoyan and Povolotsky \cite{papoyan97} pointed
out that such connections could be wrong
and used the example of the plane triangular (pt) 
lattice to illustrate their ideas.
In 1996, Vespignani, Zapperi, and Loreto \cite{vespignani96}
proposed a real space dynamical driven  
renormalization group to provide a theoretical basis for 
previous RG studies of the ASM.
In 1997, Lubek and Usadel used extensive Monte Carlo simulations 
to find that $\tau \approx 1.33$ \cite{lubek97}.

One important step in the RG approach to a lattice sandpile model is 
the exact enumeration of all possible toppling processes of sandpile
 dynamics inside a cell for RG transformation. 
PVZ and Ivashkevich  
divided the height configurations inside a RG cell into
subsets with the same number of critical height by hand.
After this procedure, for each height configuration, 
PVZ counted the possible toppling events
by hand and Ivashkevich did this by the technique of generating function.
Based on Refs. \cite{pietronero94,ves95,ivashkevich96}, 
in the present paper
we formulate a computer algorithm for RG studies of 
the BTW sandpile model on the sq, pt, and honeycomb (hc) lattices.
The advantage of our method is that we don't
need to classify all height configurations inside a RG cell 
into different subsets by hand  which is 
different for different kinds of lattices and RG cell sizes.
Therefore, it is easier for us to extend the same algorithm to different 
cell sizes and different kinds of lattices.
Another advantage of our method is that we can study larger RG cells.
For example, in \cite{pietronero94,ivashkevich96,hasty98}
and \cite{mor99}, they have only carried out $2 \times 2$ cell to one site
and 5-site cell to one site RG transformations for the sq lattice,  
and we can carry out $3 \times 3$ cell to one site RG 
transformation (RGT) for the sq lattice.

In the present paper, we consider both PVZ and Ivashkevich RGTs.
Using the PVZ reduced height RG equations,
we calculate the density of critical sites $\rho$,
the probability for one site to transfer sands to $i$ 
different nearest neighbors $p_i$, 
the avalanche exponent $\tau$ and the dynamical 
exponent $z$ by $2 \times 2$ and $3 \times 3$ cells to one site 
RGTs for the sq lattice, and 
3 sites cell to one site RGT for the pt lattice.
Using the Ivashkevich's real height RG equations,  
we calculate the critical 
height probabilities, the critical sand transfer probabilities,
the avalanche exponent $\tau$, and the dynamical
exponent $z$ by $2 \times 2$ cell to one 
site RGT for the sq lattice and 
3 sites cell to one site RGT for the pt lattice. 
We find that our calculated  height probabilities are consistent
with the exact and numerical simulations.

The outline of this paper is as follows.
In Sec. II, we introduce the reduced height RG equations.
Here, we extend PVZ's formulation for the sq lattice to a general
lattice. In Sec. III, we present a systematic computer algorithm
for the exact enumeration of all possible toppling processes of sandpile
 dynamics inside a cell for RGT.
By using this algorithm, we present the obtained fixed points of
 reduced height RG equations and the critical exponents in Sec. IV.
In Sec. V, we apply real height RG equations to 
the BTW model on sq and pt lattices.
For a different definition of RG cell for the hc lattice, in Sec. VI
we calculate the critical sand transfer probabilities and 
find that they are different from 
branching probabilities of spanning trees on the same (hc) lattice.
This result confirms Papoyan and Povolotsky's result \cite{papoyan97} 
for the pt lattice. 
Finally, we summarize our results and discuss problems for further
studies in Sec. VII. 

\section{reduced height renormalization group equations}

In this section we extend PVZ's formulation of RG transformation
\cite{pietronero94} for the sq lattice to a general
lattice $\Re$ and present a simplified version which is useful for realization
by a computer program.
Then, we propose an algorithm for counting events of RG procedure.
In the next section, we will show
that this calculation is basically consistent with that of Ref. \cite{ves95}
for a $2 \times 2$ cell of the square lattice. However, there are some mistakes
in Ref. \cite{ves95}.

As in ordinary real space RG transformations for the spin
and percolation models \cite{hu82,hu92a,hu92b,hu95}, in the RG transformation 
for the BTW sandpile model on the lattice $\Re$ with lattice constant $a$,
we first divide $\Re \equiv \Re^{(0)}$
into $g \times g$ cells, then consider the transformation from each
$g \times g$ cell into a cell with a smaller number of sites or a supersite.
 In the present paper, we consider only the later case.  
From the relationship between the cell and the supersite, 
the properties of the specified cell can be approximately 
represented by the supersite. After the first step of RG transformation,
we have a lattice, call $\Re' \equiv \Re^{(1)}$, of supersites with lattice constant $ag$. 
We can carry out the second step RG transformation on $\Re'$ to 
obtain a new lattice  $\Re''\equiv \Re^{(2)}$ with lattice constant $ag^2$
and we can continue such RG transformation to obtain a series of lattices
$\Re^{(k)}$ with lattice constant $a g^k$. 
Examples of RG cells for sq and pt lattices are shown in Fig. 1(a) and 1(b),
respectively, whose linear dimensions (i.e. $g$ values) are 2 and $\sqrt{3}$,
respectively.

Here we briefly review PVZ's RG approach to the BTW's ASM \cite{pietronero94,ves95}. 
If we consider a sandpile RG cell with $m = g^2$ sites,
 the initial height $z_i$ of each site
 $i$ inside a RG cell can take integer values 
$0, 1, \dots,$ or $q-1$, where
$q$ is the coordination number of lattice $\Re$; for example
Figs 1(a) and 1(b) correspond to the sq and the pt lattices,
with $q =$ 4 and 6, respectively. 
After adding one particle to one lattice site, say site $i$,
if the site $i$ is unstable
(i.e., $z_i \ge z_c$),  the site $i$ 
will topple and transfer particles to $e_i$ different nearest neighbors.
In general, $e_i$ can take values $1, 2, \dots,$ or $q$ after RG transformation
which is different from $e_i=q$ for the original sandpile evolution on $\Re^{(0)}$. 
As an illustration, the pictures of $e_i$ for the sq lattice are shown
in Fig. 2. The toppling rule $e_i$ can be
described by the transferring probability $
\vec{{\bf p}}=(p_{1},p_{2},...,p_{q})$
 where $p_{e_i}$ is the probability that the unstable site $i$ 
transfers particles to $e_i$ different nearest neighbors;
if $e_i$ is smaller than $q$, one should also specify which directions
the particles go into, which will be explained in more details in
the next section.
In general, the evolution of the ASM at the initial stage $\Re^{(0)}$,
is characterized by the initial height configuration set $Z$ = $\{ \vec{z} \}$
=$\{ (z_1, z_2, ..., z_m)|$ $z_i=0, 1, ...,$ or $q-1, \forall i \}$ and 
the toppling rule $\vec{e} = (e_1, e_2, ..., e_m)$ =($q$, $q$, ..., $q$).
That is $p_{1}= p_{2}=...$ $=p_{q-1}=0$
and $p_{q} =1$ for  $\vec{\bf p}$ at this stage. 
For example, for the ASM model on the original square lattice, 
the height and transferring probability
can take value $z_i = 0$, 1, 2 or 3 and $\vec{\bf p} =$
(0, 0, 0, 1), respectively. 
On the other hand, after the RG transformation, the height configuration set
and toppling rule configuration set are $Z$ and $E$, respectively,
where  $E$ = $\{ \vec{e} \}$
= $\{ (e_1, e_2, ..., e_m)|$ $e_i=1, 2, ...,$ or $q, \forall i \}$.

In this way, the transferring probability $\vec{{\bf p}}$ and 
the initial height probability $\vec{\bf n}$=$(n_0, n_1,..., n_{q-1})$,
where $n_i$ is the probability that the height of the supersite is $i$,   
are useful to characterize the properties of 
the coarse grained sandpile dynamics.
We use $\vec{{\bf p}}^{(k)}$ = $( p_1^{(k)}, p_2^{(k)}, ..., p_q^{(k)} )$
 and $\vec{{\bf n}}^{(k)}$ = $( n_0^{(k)}, n_1^{(k)}, ..., n_{q-1}^{(k)} )$
to denote the transferring probability 
and initial height probability on lattice $\Re^{(k)}$.
$(\vec{{\bf n}}^{(k)},\vec{{\bf p}}^{(k)})$
and $(\vec{{\bf n}}^{(k-1)},\vec{{\bf p}}^{(k-1)})$
are linked by the $k$-th RG transformation.
Instead of $p_{i}^{(0)}$ = 0 for $i < q$ and $p_q^{(0)}$=1 and 
$\sum_{i=0}^{q-1} n_i^{(0)} =1$
in the original level, 
the normalized condition
$\sum_{i=1}^q p^{(k)}_i=1$ and $\sum_{i=0}^{q-1} n^{(k)}_i=1$ are used in 
the coarse grained level.

In order to simplify the RG transformation,
PVZ reduced the height of a site $z_i$ to only three cases corresponding
to three classes: $h_i=0$ which is stable for $z_i < z_c-1 $, 
$h_i=1$ which is critical for
$z_i =z_c-1$, and $h_i=2$ which is unstable for $z_i \ge z_c$.  
Therefore, $Z$ will be replaced by
$H$ = $\{ \vec{h} \}$ = $\{ (h_1, h_2, ..., h_m)|$ $h_i = 0$ or 1, $\forall i \}$, 
The evolution rule is as follows:   
If  a critical site $i$ with height $h_i=1$ receives one particle,
its height is now
$h_i =2$ and site $i$ becomes unstable.
 Then, this unstable site will transfer particles to neighboring sites with 
the transferring probability $\vec{\bf p}$. After this action,
this toppled site becomes stable with $h_i=0$.
In this approach, the stable site is still stable even if this site receives 
 more than one particle. 
It means that the multi-toppling process of one site is omitted in PVZ's 
RG approaches. Since 
PVZ used height configuration $H$ and toppling rules configuration $E$ to 
describe the RG sandpile evolution,
the height probability vector $\vec{\bf n}$ can be simplified to
one parameter $\rho$, which is the probability of critical sites, and
$(\vec{{\bf n}}^{(k)},\vec{{\bf p}}^{(k)})$ becomes $(\rho^{(k)},\vec{{\bf p}}^{(k)})$.

An avalanche event specifies how a toppling process goes on.
Figure 3 shows a typical avalanche event of a sandpile evolution
on a three-site cell of the pt lattice.
The labels of sites and toppling directions are shown in Fig. 3(a).
Figure 3(b) shows a starting height configuration 
with site 1 being critical, site 2 being
critical and site 3 being stable,
i.e., $\vec{h}=(h_1, h_2, h_3)$=(1, 1, 0). This starting height configuration
appears with a probability $\rho^2 (1-\rho)$. When we add one particle to 
site 1, site 1 becomes unstable and transfers one particle to 
site 2 via directions 2 and out of cell via direction 6.
 Note that the probability of adding
 one particle to site 1 is 1/3 since there are 3 sites inside the RG cell and 
the probability of transferring particles to directions 2 and 6
from site 1 is $p_2/{C^6_2}$ since there are $C^6_2$ possible toppling
rules for transferring two particles to six directions.
After site 2 receives one particle from site 1 it becomes unstable and
transfers one particle to site 1 via direction 5, one particle to site 3 
via direction 6, and two particles out of cell via directions 1 and 2;
the probability of transferring particles to four nn sites from site 2
via directions 1, 2, 5, and 6 is $p_4/{C^6_4}$.
Since site 1 will not topple again and
site 3 is a stable site, the toppling process stops. In summary, 
the probability of the avalanche event and the relaxation time ($t$)
of Fig. 3(b) are $(1/3) \rho^2 (1-\rho) (1/15)p_2(1/15) p_4$ and
2 (we set $t_w=1$), respectively.

Under RG transformation, each three-site cell of $\Re^{(k-1)}$
transfers into one supersite of $\Re^{(k)}$
as shown in Fig. 1(b) and Fig. 3(c); a site of $\Re^{(k-1)}$
can transfer particles to six directions shown in the right-hand
side of Fig. 3(a) and a supersite of $\Re^{(k)}$ can 
transfer particles to six directions shown in the right-hand
side of Fig. 3(c), which correspond to six outgoing directions
of Fig. 1(d). In the avalanche event of Fig. 3(b), the particles
which leave the cell in directions 6, 1, and 2 contribute to
outgoing particles in directions 5, 1, and 2, respectively,
in the figure at the right-hand side of Fig. 3(c) (see also
Fig. 1(d)). Thus the avalanche event of Fig. 3(b) in $\Re^{(k-1)}$
contributes to $p^{(k)}_3$ for the supersite of $\Re^{(k)}$. 

In general, the probability of one avalanche event is determined by 
the initial height configuration,
the starting site of the avalanche, and the detail of the toppling process.
If the initial height configuration is the $j$-th element of $H$ and contains 
$m'$ critical sites, the starting site is site $\alpha$, the toppling process
consists of $a_k$ unstable sites which transfer $k$ particles to $k$ different
directions for $1 \le k \le q$, the probability of this avalanche event is 
\begin{equation}
P_E(j, \alpha, a_1, a_2, \dots, a_q)=
W_{j} W_{<\alpha>} (\frac{1}{C^q_1}p_1)^{a_1}(\frac{1}{C^q_2}p_2)^{a_2}...
(\frac{1}{C^q_q}p_q)^{a_q},
\label{pe}
\end{equation}
 where $W_{j}=\rho^{m'}(1-\rho)^{m-m'}$ and
$W_{<\alpha>}=1/m$,
$a_1, a_2,$ ..., and $a_q$
are zero or positive integer and $\sum_{i=1}^q a_i =m''$ which is 
the total number of toppled sites.
Since we don't consider multiple toppling events (i.e., one site can 
topple at most once),
$m''$ should be equal to or smaller than $m$.

To construct RGT, we should first collect all possible events on a RG cell
of $\Re^{(k-1)}$.
Consider the set of events $C_j(\alpha, a_1, a_2, \dots, a_q, t)$
which are initiated by the $j$-th initial height configuration of $H$
and starting site $\alpha$, 
have relaxation
time $t$, 
topple $a_k$ times by toppling rule with $k$ directions.
For the set $C_j(\alpha, a_1, a_2, \dots, a_q, t),$ we count
the number of events on $\Re^{(k-1)}$ which topple $i$ directions 
in point of view of site on $\Re^{(k)}$. 
This number is denoted by $B_i( j, \alpha, a_1, a_2,..., a_q,t)$.
Therefore, for $1 \le i \le q$
\begin{eqnarray}
f_{ij}&=&\sum_{\alpha} \sum_{a_1, a_2,..., a_q} \sum_t B_i( j, \alpha, a_1, a_2,..., a_q,t) 
W_{<\alpha>}
(\frac{1}{C^q_1}p_1^{(k-1)})^{a_1}(\frac{1}{C^q_2}p_2^{(k-1)})^{a_2}
\dots (\frac{1}{C^q_q}p_q^{(k-1)})^{a_q}  \nonumber \\
&=&\sum_{\alpha} \sum_{a_1, a_2,..., a_q} \sum_t B_i( j, \alpha, a_1, a_2,..., a_q,t)
P_E(j, \alpha, a_1, a_2, \dots, a_q)/W_j
\label{fij}
\end{eqnarray} 
is the summation of probability of toppling events on $\Re^{(k-1)}$ which evolve 
under the $j$-th initial configuration on $\Re^{(k-1)}$
and contribute to the $p^{(k)}_i$ on $\Re^{(k)}$.  
Note that it is possible to have toppling processes inside a cell but 
transfer nothing outside a cell. PVZ omits all of these events and Ivashkevich
et. al. \cite{ivashkevich99} simply assumed these events appear with probability $p_0$.
Now the properties of the system are fully characterized by the distribution 
$( \rho^{(k)},\vec{{\bf p}}^{(k)})$ at this scale. Then, the RG equation 
can be written as
\begin{equation}
\left(
\begin{minipage}{0.9cm} 
$p^{(k)}_1$ \\
$p^{(k)}_2$ \\
$\vdots$ \\
$p^{(k)}_q$ 
\end{minipage} 
\right)
= W_1 /N_w
\left(
\begin{minipage}{1.1cm} 
$f_{11}/F_1$ \\
$f_{21}/F_1$ \\
~~~~$\vdots$ \\
$f_{q1}/F_1$ 
\end{minipage}
        \right)
+ W_2 /N_w
\left(
\begin{minipage}{1.1cm} 
$f_{12}/F_2$ \\
$f_{22}/F_2$ \\
$\vdots$ \\
$f_{q2}/F_2$ 
\end{minipage}
        \right)
+...... 
+ W_n /N_w
\left(
\begin{minipage}{1.1cm} 
$f_{1n}/F_n$ \\
$f_{2n}/F_n$ \\
$\vdots$ \\
$f_{qn}/F_n$ 
\end{minipage}
        \right)
\label{rg1}
\end{equation}
where $n$ is the total number of configurations in $H$ which could contribute
to the right-hand side of Eq. (\ref{rg1}),
$W_j$, the weight of the $j$-th element of $H$,
 is a function of $\rho^{(k-1)}$,
$N_w$ and $F_j$ are the normalized constants of $W_{j}$
and $f_{ij}$ with $N_w=\sum_{j} W_{j}$ and $F_j=\sum_i f_{ij}$
, respectively.   
This RG equation can be well understood by the transition rate and
master equation of sandpile evolution \cite{vespignani96,ivashkevich99}.

Another important assumption is the
inflow of particles equals the flow of particles out of the system, i.e.,
in the stationary state, 
$
\frac{\partial}{\partial t} < \rho >=0
$ \cite{pietronero94,ivashkevich99}.
This implies $\rho^{(k-1)}$ is a function of 
$p_j^{(k-1)}$ for $j$ from 1 to $q$ and can be written as
\begin{equation}
\rho^{(k-1)} = \frac{1}{\sum_i i \times  p_i^{(k-1)}}.
\label{rg2}
\end{equation}
Replacing $\rho^{(k-1)}$ in $W_j$ and $N_w$ of Eq. (\ref{rg1}) by Eq. (\ref{rg2}),
$ p_i^{(k)}$ is a pure function of $p_j^{(k-1)}$ for $j$ from 1 to $q$.
From RG assumption,
at $k \rightarrow \infty$, the critical transferring probability $\vec{{\bf p}}^*$
= $(p_1^*, p_2^*, ..., p_q^*)$ and the critical density $\rho^*$ can be obtained.

\section{a simple algorithm for counting events}

In lattices $\Re^{(k)}$ for $k \ge 1$ obtained after RG transformations,
one site can transfer particles to one, two, ..., or $q$
different directions with probabilities $p_1$, $p_2$, ..., 
and $p_q$, respectively. 
There are totally $q$ different directions which can be labeled by
$d_1$, $d_2$, ..., and $d_q$.
In a more precise description,
a site can transfer one particle or nothing to a specified direction $d_i$ 
which is denoted by the variable $r_{d_i}$ with values
$r_{d_i} = 1$ or 0, respectively. 
Therefore, the toppling rules of sites  are exactly determined by
the vector ${\vec{\bf r}} = (r_{d_1},r_{d_2}, ..., r_{d_q})$.
Since one unstable site sends nothing to its neighbors is omitted
in PVZ RG transformation,
$(r_{d_1}, r_{d_2}, ..., r_{d_q})$ = $( 0, 0, ..., 0)={\vec 0}$ is forbidden.
For a given ${\vec{\bf r}}$, the site will send 
$j=\sum_{i=1}^{q} r_{d_i}$ particles
to $j$  different directions.
If we divide the toppling rule configuration set $\Im$ = $\{(r_{d_1}, 
r_{d_2}, ...,r_{d_q}) |
r_{d_i} = 0$ or 1 , $\forall i$ and $\sum_{i=1}^q r_{d_i} \ne 0 \}$ 
into subsets $\Im_j$ = $\{(r_{d_1}, r_{d_2}, ...,r_{d_q}) |
r_{d_i} = 0$ or 1 , $\forall i$ and $\sum_{i=1}^q r_{d_i} =j \}$
according to the value $j$,
$q$ subsets can be obtained. 
And it is straightforward to show that the number of elements of the $j$-th subset $\Im_j$ 
and $\Im$ 
are $C^q_j$ and $\sum_{j=1}^{q} C^q_j=2^q-1$, respectively.

Now we define $Pro.( {\vec{\bf r}} )$ as the probability that an unstable site transfers
particles with a given toppling rule ${\vec{\bf r}}$. 
The relation $p_j = \sum_{{\vec{\bf r}} \in \Im_j} Pro.( {\vec{\bf r}} )$
can be obtained. On the other hand, we assume that the probability $Pro.( {\vec{\bf r}} )$
 is the same for all ${\vec{\bf r}} \in$
 $\Im_j$ because of the isotropy of lattice $\Re$.  
Therefore, we conclude that $Pro.( {\vec{\bf r}} ) = p_i/C^q_i$ for ${\vec{\bf r}} \in$
$\Im_i$. For example,
$Pro.({\vec{\bf r}} = (1,0,...,0))$ = $Pro.({\vec{\bf r}} = (0,1,...,0))$
 =...= $Pro.({\vec{\bf r}} = (0,0,...,1))$
= $p_1/q$ and $Pro.({\vec{\bf r}} = (1,1,...,0))$ = $p_2/C^q_2$.
In this way ${\bf R}$=$\{( {\vec{\bf r}}_1, {\vec{\bf r}}_2, ..., 
{\vec{\bf r}}_m )| \vec{{\bf r}}_i \in \Im, \forall i \}$ can be used to
represent the toppling rule configuration of an RG cell;
${\bf R}$ contains more information
than $E=\{\vec{e}\}$ introduced in Sec. II.

In Sec. II, we presented PVZ approach to construct RG transformation of 
Eq. (\ref{rg1}) by using avalanche events and we used Fig. 3 as an 
illustrative example. In this section, we will use the idea of 
toppling rule configuration ${\bf R}$ and a
computer algorithm to calculate all terms which contribute to the 
right-hand side of Eq. (\ref{rg1}).
Consider a RG cell of $m$ sites, whose initial high configuration is
$( h_1, h_2, ..., h_m )$ and the toppling rule configuration is 
$( {\vec{\bf r}}_1, {\vec{\bf r}}_2, ..., {\vec{\bf r}}_m )$.
If we add one particle to site $\alpha$,
 the sandpile evolution starts and finally
evolves into one state where all new $h_i$ are stable.
In order to count all possible avalanche events, we prepare a Fortran
computer program SOCRG to generate such events. In SOCRG, we consider
three sub-configurations sets and use $(m+2)$ do-loops to
generate such sub-configurations:
\begin{enumerate} 
\item  The initial height configuration $H$: There are two possible states
for each site i.e., $h_i=0$ or 1. 
Therefore, totally, $2^m$ possible configurations are considered.
However, some of these $2^m$ configurations will not induce sandpile evolution,
for example, $h_i = 0$ for all $i$.
In SOCRG, we use the first do-loop to generate all possible configurations
in $H$. 

\item  The starting point set of sandpile dynamics
 $\Psi = \{\alpha | \alpha = 1, 2, ...,$ or m $\}$, 
where $\alpha$ denotes the site on which we add one particle:
There are $m$ possible positions to add one particle to a RG cell 
with $m$ sites. In SOCRG, we use the second do-loop to generate
all possible site $\alpha$ in $\Psi$. If the chosen site $\alpha$ is
not a critical site, SOCRG goes to the next site.

\item  The toppling rule configuration ${\bf R}$:
There are $2^q - 1$ possible toppling rules for each site. 
Therefore, there are ${(2^q - 1)}^{m}$ possible toppling configurations 
for an $m$-site cell.  In SOCRG, we use the third to the $(m+2)$-th
do-loops to generate such toppling configurations. 

\end{enumerate}

For a specified configuration with $\vec{h}^s$, $\alpha^s$, 
$(\vec{\bf r}^s_1, \vec{\bf r}^s_2, \dots, \vec{\bf r}^s_m)$
 generated by these $(m+2)$ do-loops,
we can calculate the toppled vector
$\vec{O}=( O_1, O_2, \dots, O_m)$ where $O_i$=1 or 0
depending on whether site $i$ topples or it doesn't
during the sandpile evolution. For example,
Fig. 3 shows an event on a three-sites RG cell of the triangle lattice.
In this example, $(h_1^s, h_2^s, h_3^s )$ obtained from the first do-loop, 
$\alpha^s$ obtained from the second do-loop,
${\vec{\bf r}}^s_1$ obtained from the third do-loop, 
and ${\vec{\bf r}}^s_2$ obtained from the 4-th do-loop, 
are (1, 1, 0), 1, (0, 1, 0, 0, 0, 1), and
(1, 1 ,0, 0, 1, 1), respectively. Then, $\vec{O}=(1, 1, 0)$ is obtained.
The weight of this height configuration is $\rho^2 (1-\rho)$. 
The probability of this event is 
$ 1/3 \rho^2 (1-\rho) Pro.(\vec{\bf r}^s_1=(0, 1, 0, 0, 0, 1)) 
Pro.(\vec{\bf r}^s_2=(1, 1 ,0, 0, 1, 1))$ 
and the relaxation time $t$ = 2.
Finally, this event is classified as ${\vec{\bf r}}$ 
= $(1, 1, 0, 0, 1, 0)$ for the supersite of Fig. 3(c).
In this case, since site 3 does not topple ($O_3=0$).
the probability of the toppling process depends only on 
 $\vec{h}^s$, $\alpha^s$, ${\vec{\bf r}}^s_1$, and ${\vec{\bf r}}^s_2$.
The do-loop used to generate ${\vec{\bf r}}^s_3$ can be skipped
quickly (see below).
 
In general, consider the $i_1$-th element of the first do-loop $\vec{h}^{i_1}$,
the $i_2$-th element of the second do-loop $\alpha^{i_2}$, and the $j_k$-th 
element of the $(k+2)$-th do-loop $\vec{\bf r}^{j_k}$ for $1 \le k \le m$;
the combination of such elements is
denoted by the $D(i_1, i_2, j_1, \dots, j_k, \dots, j_m)$, which is
a configuration of
the set consists of $H$, $\Psi$, and ${\bf R}$.
This configuration will generate a toppling event with probability
\begin{equation}
P_P(\vec{h}^{i_1}, \alpha^{i_2}, \{\vec{\bf r}^{j_k}\})=W_{\vec{h}^{i_1}}W_{<\alpha^{i_2}>} 
\prod_{k=1}^{m} (Pro.(\vec{\bf r}^{j_k}))^{O_{k}}.
\label{pp} 
\end{equation}
Since $O_k=0$ means that site $k$ is not involved in the toppling process,
any choice of $\vec{\bf r}$ in the $(k+2)$-th loop will correspond
the same event represented by the configuration 
$D(i_1, i_2, j_1, \dots, j_k, \dots, j_m)$.
Define a set $\aleph$ which consists of 
the configurations: $D(i_1, i_2, J_1, \dots, J_k, \dots, J_m)$ 
with $J_k = j_k$ for $O_k=1$ or $J$ for $O_k=0$, where 
$J$ is an integer and $1 \le J \le 2^{q-1}$.
From $\vec{O}$, we can calculate the total number of toppled sites:
$m''=O_1+O_2+\dots+O_m$.
There are $(2^q-1)^{m-m''}$ elements in $\aleph$ and
every such element of $\aleph$ corresponds to the same event which
 is generated by the  configuration
$D(i_1, i_2, j_1, \dots, j_k, \dots, j_m)$. In SOCRG, the do-loops
corresponding to sites with $O_k=0$ are passed through quickly to
save the computing time and the following technique is used.

Define an $m$-dimensional array: $V(j_1, j_2, \dots, j_m)$,
where $1 \le j_k \le 2^{q-1}$ for $1 \le k \le m$.
Immediately after a new height configuration is chosen by the first do-loop
(say the $i_1$-th step) and a new starting point is chosen by the second do-loop
(say the $i_2$-th step), all elements of the array $V$ are set to 1.
For each $D(i_1, i_2, j_1, \dots, j_k, \dots, j_m)$ configuration 
generated in the third to the $(m+2)$-th do-loop,
we first check the value of $V(j_1, j_2, \dots, j_m)$.
If $V(j_1, j_2, \dots, j_m)$=1, we record this event and obtain a set $\aleph$.
Then, we set $V(J_1, J_2, \dots, J_m)=0$ for $D(i_1, i_2, J_1, J_2, \dots, J_m)$
in $\aleph$. If $V(j_1, j_2, \dots, j_m)$=0, we skip this step to the next step 
under the third do-loop to the $(m+2)$-th do-loop.
In this way, the do-loops corresponding to sites with $O_k=0$ 
can be passed through quickly.
Repeat above procedure for different combinations of 
height configuration and starting point and we finish the 
calculation of all do-loops.

Basically, the form of Eq. (\ref{pp}) can be transferred to the form of Eq. (\ref{pe}).
Therefore, we can construct the RG Eqs. (\ref{rg1}) and (\ref{rg2}).
In order to test this algorithm, we calculate RGT for a
$2 \times 2$ cell to one site on square lattice,
which is shown in Figs. 1(a) and (c)
and has been done in details by Vespignani, Zapperi, and Pietronero (VZP)
 \cite{ves95}.
Define $A_{i}(k, a_1, a_2, a_3, a_4)$ = $\sum_{j} 
\sum_{\alpha} \sum_t B_i(j, \alpha, a_1, a_2, a_3, a_4, t)(\frac{1}{C^4_1})^{a_1}
(\frac{1}{C^4_2})^{a_2}
(\frac{1}{C^4_3})^{a_3}
(\frac{1}{C^4_4})^{a_4}
 \times \delta (k,\phi (j))$
where $\phi (j)$ is the number of critical sites of the $j$-th configuration of heights.
We find that almost all of our calculated values of $A_{i}(k, a_1, a_2, a_3, a_4)$
  are the same as those 
which appear in the appendix of \cite{ves95},
except $A_3(3,0,1,1,1)=0.749997$, $A_2(4,0,4,0,0)=0.1772839$, and $A_4(4,2,0,1,1)=1.460250$
in \cite{ves95}. Our values of $A_3(3,0,1,1,1)$
, $A_2(4,0,4,0,0)$, and $A_4(4,2,0,1,1)$ are 1.750000, 0.1172839, and 1.406250, respectively.
We find that $A_2(4,0,4,0,0)$ and $A_4(4,2,0,1,1)$ in Ref. \cite{ves95} are 
only slightly different
from our values. They might be typos in \cite{ves95}.
However, there is an obvious difference  between our value and VZP's
 value for $A_3(3,0,1,1,1)$
We believe that VZP's  value is wrong since all other 240 terms of VZP
 are exactly the same as ours
and VZP calculated $A_{k}(i, a_1, a_2, a_3, a_4)$ by hand and
 we calculate $A_{k}(i, a_1, a_2, a_3, a_4)$ by a systematic algorithm.

\section{Calculations of critical exponents}

In the following calculations we use the RG approach in the 
form of Eqs. (\ref{rg1}) and (\ref{rg2}).
First, we consider the RG calculation 
from the small scale transformation for square and triangle
lattices which are shown in Fig. 1.
All bonds outgoing from a given cell into another are included
into one renormalized bond of the supersite, 
as is shown in Figs. 1(c) and 1(d).
It is one of the simplest choices where supersites
 of Figs 1(c) and 1(d) contain just 
a few sites of Fig 1(a) and Fig. 1(b), respectively. 
A block of sites in $\Re^{(k-1)}$ (i.e. an RG cell)
is replaced by a site of $\Re^{(k)}$.
Here g is equal to 2 and $\sqrt{3}$ for 
square and triangle lattices, respectively.
A more complex choice for the square lattice
 can be found in \cite{mor99} where
one supersite contains five sites.  
In the present paper, for the square lattice
we will extend this study to the case where one RG cell
contains nine sites. This is the largest RG cell which has been
considered for the RG approach to the BTW sandpile model.

By using the algorithm of section III, we enumerate all  
possible toppling events for two kinds of 
cells shown in Fig. 1. However, as in \cite{pietronero94,ves95}
we drop those events in which the number of toppled sites 
are smaller than $g_t =$2; here it is assumed that such events
contribute to $p_0$ which has been discussed in Sec. II.
In other words, these events are assumed to transfer no particle 
to nearest neighbor cells. Of course, this is an approximation.
From the procedures stated above, we can express
 $W_j$ and $f_{ij}$ of Eq. (\ref{rg1}) in term of 
$\vec{{\bf p}}^{(k-1)}$ 
and $\rho^{(k-1)}$. The normalized factors 
$W_n$ and $F_i$ can also be calculated.    
By using initial values of $\vec{{\bf p}}^{(0)}$ 
and $\rho^{(0)}$ to iterate Eqs. (\ref{rg1}) and (\ref{rg2}) 
until $k \rightarrow \infty$, we can obtain the fixed point 
of $ \vec{{\bf p}}$ and $\rho$ shown in Table I. 
Note that our results for $2 \times 2$ cell 
to one site RG transformation 
for the square lattice are slightly different from those of 
Refs. \cite{pietronero94,ves95}. This is due to the difference in 
values of $A_j(i,a_1,a_2,a_3,a_4)$ discussed in the Sec. III. 
 For each lattice, we choose some different starting points
 of $\rho^{(0)}$  and $\vec{{\bf p}}^{(0)}$. We find that 
all starting points evolve into the same fixed point which is 
shown in Fig. 4. This shows that there is only one 
attractive fixed point of RG equation for square and triangle lattices
in this RG transformation.

In \cite{pietronero94}, the critical exponent 
of the avalanche distribution
$\tau$ in $P(s) \sim s^{-\tau}$ is calculated at the fixed point of
the RG transformation 
through the fixed point parameters $\rho^*$ and $\vec{{\bf p}}^{*}$.
First, the probability that one avalanche occurs on $\Re^{(k-1)}$ but doesn't
occur on $\Re^{(k)}$ can be written as 
\begin{equation}
K=\sum_{i=1}^q p_i^* (1-\rho^*)^i.
\label{rgk1}
\end{equation}
In addition, the probability $K$ can also be understood as
 the probability that the linear dimension of the avalanche, $l$,
 is larger than $a g^{k-1}$ 
 ( the lattice constant of $\Re^{(k-1)}$ ) and smaller than
$a g^k$ (the lattice constant of $\Re^{(k)}$). 
Thus $P(s)ds = P(l)dl$ $\sim s^{-\tau}ds$ =$l^{1-2\tau} dl$, 
where the toppling area $s \sim l^2$, and we have  
\begin{equation}
K=\frac{\int_{a g^{k-1}}^{a g^k} P(l)dl}{\int_{a g^{k-1}}^{\infty} P(l)dl}
=1-g^{2(1-\tau)}.
\label{rgk2}
\end{equation}
also holds.
Using Eqs. (\ref{rgk1}) and (\ref{rgk2}), 
we obtain $\tau$ which is listed in the part (A) of Table II. 
We find that our $\tau$ for square and triangle lattices are close to 
the numerical simulations \cite{lubek97,manna91} for the square lattice.

Another independent critical exponent is the dynamical exponent $z$.
From the scaling laws at the fixed point we know that
 the average time of a dynamical
process scales with the linear length as 
$<t> \sim l^z.$
Therefore, the time scale $t_{k}$
of a relaxation event on lattice $\Re^{(k)}$ and $t_{k-1}$ 
on the lattice $\Re^{(k-1)}$ 
are related by the relation 
$
{t_{k}}/{t_{k-1}}=({a g^{(k)}}/{a g^{(k-1)}})^z=g^z.
$
On the other hand, the time scale $t_{k}$ can be obtained as a function of the
time scale $t_{k-1}$ from the RG equations. 
The relation is given by $t_{k}=<t^{(k-1)}> t_{k-1}$,
and 
\begin{eqnarray}
<t^{(k-1)}>&=& \sum_{i, j, \alpha, a_1, a_2,..., a_q, t_{k-1}} \frac{W_{i}}{N_w} 
\frac{W_{\alpha} B_j(i, \alpha, a_1, a_2,..., a_q, t_{k-1})}
{F_i} \nonumber \\
&&\times 
(\frac{1}{C^q_1}p_1^{(k-1)})^{a_1}(\frac{1}{C^q_2}p_2^{(k-1)})^{a_2}
...(\frac{1}{C^q_q}p_q^{(k-1)})^{a_q} (t_{k-1})
\label{rgt1}
\end{eqnarray}
where $<t^{(k-1)}>$ is the average number of subprocesses on $\Re^{(k-1)}$
needed to have a relaxation process on $\Re^{(k)}$.
By inserting the fixed point parameters into the calculation of 
$<t>=\lim_{k \rightarrow \infty} <t^{(k)}>$,
we obtain the following result for the dynamical exponent \cite{pietronero94}
\begin{equation}
z=\frac{\ln <t>}{\ln (g)}.
\label{rgt2}
\end{equation}
Using Eqs. (\ref{rgt1}) and (\ref{rgt2}),
 we obtain $z$ which is listed in the 
part (A) of Table II.
We find that our result $z=1.284$ for the triangle lattice is not far from the
theoretical prediction value $z=1.25$ \cite{majumdar92}. 
If the universality is valid for square and triangle
lattices, the obtained $z$ must be the same both lattices.
In our RG calculations, we find $z$ for the square lattice
has a larger deviation from the theoretical predicted value:
$z=1.25$.

We also consider $3 \times 3$ cell for the square lattice, which
is shown in Fig. 5. 
Here $g_t$ = 3.
The critical density of sites and transferring probabilities are
shown in Table I.
We find that there are some difference between the $\rho^*$ and $p_i^*$ 
in $2 \times 2$ cell to one site and $3 \times 3$ cell to one site RG 
transformations.  The exponents of $\tau$ and $z$ shown
 in Table II have larger deviations 
 from the simulation and theoretical prediction results 
than the results obtained 
from $2 \times 2$ cell.

\section{real height Renormalization group equations}

In the above study, the height configuration $Z$ is simply characterized 
by $H$. In this section, the real height configuration 
$Z$ is used to build the RG equation.
Instead of ($\rho$, $\vec{{\bf p}}$) in above calculations,  
($\vec{{\bf n}}$, $\vec{{\bf p}}$) is used
in this section, where $\vec{{\bf n}}=(n_0, n_1,..., n_{q-1})$ with 
$n_0+  n_1+ ... +n_{q-1} =1$ and $n_i$ is the probability 
that the height is $i$. 
Therefore the $W_i$ of Eq. (1) is 
replaced by $W_i = \Pi_{j=1}^{m} n_{z_j}$. 
And the steady state equation of Eq. (\ref{rg2}) is now 
characterized by the following conditions: 
$\dot{n}_0 =\dot{n}_1 = ... =\dot{n}_{q-1}=0$ \cite{ivashkevich96}.
Therefore, the relationships between concentration 
of height $\vec{{\bf n}}^{(k)}$ 
and transferring probability $\vec{{\bf p}}^{(k)}$ 
at the stationary state  is: 
\begin{equation}
 n_{i-1}^{(k)} = (\sum_{j=1}^i p_{q-j}^{(k)})/\overline{p^{(k)}}
\label{rgn1}
\end{equation}
where $\overline{p^{(k)}} = \Pi_{i=1}^q i \times p_i^{(k)}$ 
is the average number of particles sent from one site to 
other sites.

By using the computer algorithm in section III, we count all events over
$Z \otimes \Psi \otimes R$. Here, for height configuration $( z_1, z_2,..., z_m )$,
there are $q$ possible states for each site, i.e., $z_i$ = 0, 1,..., or $q-1$.
Totally, $q^m$ possible configurations are considered. Therefore,
the number of events with real height $Z$ is much larger than that with $H$. 
According to Eqs. (\ref{rg1}) and (\ref{rgn1}), the $p_i^*$ and $n_i^*$ can be obtained
after repeated iterations. 
In Table III, we compare the critical height probabilities with 
the exact \cite{priezzhev94} and numerical
results. The numerical results
are obtained from simulations on $1000 \times 1000$
 sq and pt lattices and in each case $10^6$ configurations are generated
to obtain the data.
We also list the results of RG calculations reported in
Refs. \cite{ivashkevich96,papoyan97}.
We find that our RG fixed point is very close to previous 
RG calculations \cite{ivashkevich96,papoyan97}.
It means that our RG calculation is reliable.
We also find that the RG results of the
critical height probabilities $n^*_i$ have the same behavior as 
the numerical or exact results.

For critical exponent $\tau$, Eq. (\ref{rgk2}) is still 
useful but Eq. (\ref{rgk1}) should be
revised to satisfy the definition of $\vec{{\bf n}}$. The new equation is 
\begin{equation}
K = \sum_{i=1}^q p_i^* (1- n_q^*)^i 
\label{rgk3}
\end{equation}
where $\rho^*$ in Eq. (\ref{rgk1}) has to be replaced by $n_q^*$.
By Eqs. (\ref{rgk2}) and (\ref{rgk3}), the critical exponent 
$\tau$ is obtained for sq and triangle
lattice which are shown in the part (B) of Tab. II.
We find that the value of $\tau$ is very close to previous calculation for
sq and triangle lattice. It shows again that our computer algorithm and
calculation are equivalent to Ivashkevich's algorithm.
In Refs. \cite{ivashkevich96,papoyan97}, they didn't calculate the dynamical
exponent $z$. Here, by Eqs. (\ref{rgt1}) and (\ref{rgt2}), we calculate this quality 
which is shown in part (B) of Tab. II.
We find that the obtained $z$ on triangle lattice is far from
 the value of exact and numerical results for the square lattice.

\section{Calculations for the honeycomb lattice}

There is no overlap of sites between RG cells in the Figs.1 and 5.
In other words, from these two figures, if one site belongs to one specified cell,
this site does not belong to other cells.
 Due to this property of cells, Eqs. (\ref{rgk1})-(\ref{rgk3})
can be used to calculate critical exponents.
Consider the RG transformation cell for the honeycomb (hc) lattice shown
in Fig. 6.
This kind of RG cell allows one site to belong to two different RG cells.
If we use the cell transformation in Fig. 6(a) and 6(b),
we can still  obtain the fixed point
of $\rho$ and $p_i$. However, it seems that it is inappropriate
to get the critical exponents $\tau$ and $z$
for the hc lattic by Eqs. (\ref{rgk1})-(\ref{rgk3}).

The algorithm in Sec. III is still useful for the calculation of
transferring probabilities for the 
honeycomb lattice. First, for the reduced height RG equation, 
we count all possible toppling events and drop those events with $g_t =3$.
Then, using Eqs. (\ref{rg1}) and (\ref{rg2}) to obtain $\rho^*$ and $p_i^*$ which
are shown in Table I. For real height RG equation, the obtained
$n_i^*$ and $p_i^*$ are shown in Table III and IV, respectively.
We find that the height probabilities $n_i$ are quite consistent with the 
numerical simulations on $1000 \times 1000$ honeycomb lattice, where
$10^6$ configurations are generated to obtain numerical data.

The branching probability of a spanning tree \cite{dhar90}
$\hat{p}_k$ is the probability for 
any site of a random spanning tree to have a coordination number $k$.
The Green functions of the Laplace equation for the square, triangle, and
honeycomb lattices are well known \cite{rwbook}. 
For an infinite lattice, $G$ depends 
only on the difference of site coordinates rather than their values.
Here we show the Green function for the honeycomb lattice 
 as an example:
\begin{equation}
 G(r_1-r_2)= \int_0^{2\pi} \int_{0}^{2 \pi}\frac{1}{2} 
\frac{f(x_1,x_2,y_1,y_2,\alpha,\beta)}
{1-\frac{4}{9}(4 cos^2 \alpha  4 cos \alpha cos \beta )}
 \frac{d \alpha }{ 2 \pi} 
\frac{d \beta}{2 \pi}, 
\label{green1}
\end{equation}
where 
\begin{eqnarray}
f(x_1,x_2,y_1,y_2,\alpha,\beta)&=&cos(\alpha (x_1-x_2)\beta (y_1-y_2))
\frac{2}{3} cos (\alpha) 
cos(\alpha (x_1-x_2)\nonumber \\
  &&\beta (y_1-y_2))\frac{1}{3} 
cos(\alpha (x_1-x_2)\beta (y_1-y_2) \beta) -1.
\end{eqnarray}
Following the same procedure as Ref. \cite{manna92}, 
we obtain the branching 
probability of a spanning tree $\hat{p}_k$ for the
honeycomb lattice.
The values of $\hat{p}_k$ for sq and pt lattices \cite{papoyan97,manna92}
 are also listed in Table IV. 
In Table IV, we compare the RG transferring probabilities $p_i^*$ 
with the branching
probabilities of spanning tree $\hat{p}_i$. We find that they have the same
behavior on the square lattice. However, they are not on the triangle and
honeycomb lattices. Thus, we can conclude that the hypothesis about the 
coincidence of $p_i$ and $\hat{p}_i$ 
proposed in \cite{ivashkevich96} is not valid.

\section{summary and discussion}

In this paper we use a computer algorithm to calculate 
the effective toppling events
for two kinds of RG equations. 
 We find that the values of critical exponents $\tau$ and $z$
for $2 \times 2$ cell to one site RG transformation for the sq lattice 
are closer to conjectured exact values and simulation
values than those for  $3 \times 3$ cell to one site RG transformation.
There are two possible reasons. The first reason is that we don't consider
the multiple toppling events in the RG calculations.
The errors arising from the multiple toppling events are larger for $3 \times 3$ cell.
Therefore, when the cell size increases, we should consider the multiple toppling events
in order to get accurate  values of $\tau$ and $z$. This can not be done easily
in exact enumeration approach to RG transformations, but can be done by Monte
Carlo RG calculations. Another possible reason is that the critical exponent
$\tau$ is not well defined for the BTW sandpile model
and therefore RGT can not be used to obtain the critical exponent.

For the PVZ approach, the critical density of sites, $\rho^*$, is
equivalent to height probabilities, $n_q^*$, in the Ivashkevich approach.
If we compare the value of $\rho^*$ with the numerical or exact $n_q^*$,
we find that $\rho^*$ is larger than $n_q^*$.
This is related to the rule that in PVZ approach the stable site will not topple
even the stable site receive more than one particles.
Therefore, the obtained $\rho^*$ must be larger to compensate the loss of
stable sites which have the potential of toppling. 
Again, when the cell size is larger, the compensation effect is also larger.
For example, $\rho^*$ obtained from $3 \times 3$ cell of the square lattice
is larger than that obtained from the
$2 \times 2$.
There is no such kind of problem in Ivashkevich's approach.
We find the $n_i^*$ is consistent with the numerical simulations and exact
results.

In summary,  it is worthwhile to consider
 the multiple toppling events  in the real height RG treatment
in order to answer the question discussed above.
However, it is hard to carry out exact enumerations for larger cell sizes.
In the next step, we plan to use Monte Carlo simulations to
construct RG transformations with large cells and
include multiple toppling events.

\begin{acknowledgments}
We thank D. Dhar and V. B. Priezzhev for useful discussion and 
A. M. Povolotsky for a critical reading of the paper.
This work was supported by the National Science
Council of the Republic of China (Taiwan) under Grant No.
NSC 90-2112-M-001-074.
\end{acknowledgments}

\begin{figure}

\caption{(a). and (b). Transformation from a cell to a site for the RG
transformation.  Here (a) and (b) are for square lattice with 
$2 \times 2$ cells and for the plane triangular lattice with 
$\sqrt{3} \times \sqrt{3}$ cells, respectively.
Blocks of sites of the coarse grained lattices $\Re^{(k-1)}$
become supersites sites on the renormalized lattice $\Re^{(k)}$.
(c). and (d). 
We show that the directions outgoing from the blocks, which
are encircled by rectangle corner roundness, are coupled 
to the directions of the lattice at the next scale.}  
\label{fig1}
\end{figure}

\begin{figure}

\caption{ The picture of toppling rules for a site of the
 square lattice.  The $e_i$ denotes one unstable
site $i$ will transfer $e_i$ particles to $e_i$ different 
nearest neighbors. 
For each $e_i$, there are $C^q_{e_i}$ possible toppling rules.
}
\label{fig2}
\end{figure}

\begin{figure}

\caption{
A toppling process on a three-site cell of the plane triangular lattice. 
Open dots represent stable sites, 
black dots represent critical sites, and encircled black
dots represent unstable site.
(a) The label of sites inside a cell and the label of  
toppling directions of an unstable site.
(b) A series of toppling process from left to right.
For site 1 and 2, their toppling rule $\vec{\bf r}_1$ and $\vec{\bf r}_2$ are
(0, 1, 0, 0, 0, 1) and (1, 1 ,0, 0, 1, 1), respectively.
(c) After the RG transformation, the three-site cell of (a)
    the toppling processes of (b) are represented  a supersite and
    the toppling rule $\vec{\bf r}$ = (1, 1, 0, 0, 1, 0) on the
    supersite.
Note that the toppling directions in (c) can be obtained from (b) by
rotating $30^{\circ}$ counterclockwise. }
\label{fig3}
\end{figure}

\begin{figure}
\caption{Iteration results for different initial values of
$\vec{\bf p}^{(0)}$ and $\rho^{(0)}$. 
The solid line with
 symbol $\circ$, $\Box$, $\bigtriangleup$, and $\bigtriangledown$
 represent the $2 \times 2$ cell transformation on square lattice with
initial values $\vec{\bf p}^{(0)}$ = (0, 0, 0, 1), (0, 0, 0, 1),
($\frac{1}{4}$, $\frac{1}{4}$, $\frac{1}{4}$, $\frac{1}{4}$), and
($\frac{1}{4}$, $\frac{1}{4}$, $\frac{1}{4}$, $\frac{1}{4}$) and 
$\rho^{(0)}$ = 0.1, 0.8, 0.1, and 0.8, respectively.
The dashed line with
 symbol $\circ$, $\Box$, $\bigtriangleup$, and $\bigtriangledown$
 represent the 3 sites cell transformation on triangle lattice with
initial values $\vec{\bf p}^{(0)}$ = (0, 0, 0, 0, 0, 1), (0, 0, 0, 0, 0, 1),
($\frac{1}{6}$, $\frac{1}{6}$, $\frac{1}{6}$, $\frac{1}{6}$, 
$\frac{1}{6}$, $\frac{1}{6}$), and
 and ($\frac{1}{6}$, $\frac{1}{6}$, $\frac{1}{6}$, $\frac{1}{6}$, 
$\frac{1}{6}$, $\frac{1}{6}$)
$\rho^{(0)}$ = 0.1, 0.8, 0.1, and 0.8, respectively.
(a). A plot of the $\rho^{(k)}$ against the $k$th iteration number $k$.
(b). A plot of the 
$|\vec{\bf p}^{(k)}|^2$ = $(p_1^{(k)})^2+(p_2^{(k)})^2+ 
 ...+(p_q^{(k)})^2$ against the $k$th iteration number $k$.
}
\label{fig4}
\end{figure}

\begin{figure}
\caption{Transformation from a cell with 9 sites on square lattice.
(a). This shows the transformation from a cell to a site. 
(b). We show that the directions outgoing from the blocks, which
are encircled by rectangle corner roundness, are coupled 
to the directions of the lattice at the next scale.}
\label{fig5}
\end{figure}

\begin{figure}

\caption{
Transformation from a cell with 6 sites on honeycomb lattice
(a). This shows a transformation from a cell to a site. 
(b). We show that the directions outgoing from the blocks, which
are encircled by rectangle corner roundness, are coupled 
to the directions of the lattice at the next scale.}
\label{fig6}
\end{figure}

\begin{table}
\caption{Critical density and transferring probabilities at the RG fixed point.}
\begin{tabular}{cccccccc}          

         &$\rho^*$ &  $p_1^*$  &  $p_2^*$  & $p_3^*$  &  $p_4^*$ & $p_5^*$ & $p_6^*$   \\
\hline \hline
sq [2 $\times$ 2$]^{(a)}$  & 0.515  &   0.327   & 0.437     &  0.205    &   0.031  & ----- & -----  
\\  \hline
sq [2 $\times$ 2$]^{(b)}$    & 0.468  &   0.240   & 0.442     &  0.261    &   0.057  & ----- & -----  
\\  \hline
sq [3 $\times$ 3$]^{(a)}$ & 0.663 &   0.572  &   0.353  &  0.070  &    0.005 & -----  & ----- 
\\ \hline \hline
pt [3 sites$]^{(a)}$ & 0.214     &   0.000002   & 0.0005     &  0.040    &   0.314  & 0.582 & 0.073 
\\ \hline \hline
hc [6 sites$]^{(a)}$ & 0.763   &   0.702  &   0.285  &  0.013  &  ----- & -----  & ----- 
\\ 
\end{tabular}
(a) : this work based on PVZ approach\\
(b) : PVZ 1994 \cite{pietronero94}
\label{tab1}
\end{table}

\begin{table}
\caption{Avalanche exponent $\tau$ and dynamical exponent $z$ for
 square and plane triangular lattices.}

\begin{tabular}{ccccc}    
  & $\tau$ (sq) & $z$ (sq) & $\tau$ (pt) & $z$ (pt) \\ \hline 
sq [$2 \times 2$]$^{(a)}$ $(A)$      &   1.243   &  1.147  & -----  & -----   \\  \hline
sq [$2 \times 2$]$^{(b)}$ $(A)$      &   1.253  &   1.168  &  -----  & -----  \\ \hline
sq [5 sites]$^{(c)}$ $(A)$      &   1.235  &   1.236  &  -----  & -----  \\ \hline 
sq [$3 \times 3$]$^{(a)}$ $(A)$      & 1.122     & 1.082 &   -----   & -----       \\  \hline
pt [3 sites]$^{(a)}$ $(A)$       &   -----  &   ----- &   1.363   &  1.284  \\ \hline \hline 
sq [$2 \times 2$]$^{(a)}$  $(B)$      &   1.248  &   1.150 &   -----   &  -----  \\ \hline
sq [$2 \times 2$]$^{(d)}$  $(B)$      &   1.248  &   ----- &   -----   &  ----- \\ \hline 
pt [3 sites]$^{(a)}$  $(B)$      &   -----   & -----&   1.367   &  1.433 \\  \hline
pt [3 sites]$^{(e)}$   $(B)$     &   -----   & -----&   1.367   &  -----  \\  \hline 
\hline 
Simulation   &   1.33$^{(f)}$   &  1.254$^{(i)}$  & -----  & -----  \\  \hline 
Prediction   &   1.25$^{(g)}$ and 1.2$^{(h)}$  &  1.25 $^{(i)}$  & -----  & ----- \\
\end{tabular}
(A): reduced parameter RG calculations \\
(B): real parameter RG calculations \\
(a): this work     \\
(b): PVZ 1994 \cite{pietronero94}  \\
(c): Moreno, Zapperi, and Pacheco 1999 \cite{mor99} \\
(d): Ivashkevich 1996 \cite{ivashkevich96} \\
(e): Papoyan and Povolotsky 1997 \cite{papoyan97} \\
(f): Lubek and Usadel 1997 \cite{lubek97} \\
(g): Priezzhev, Ktitarev, and Ivashkevich 1996 \cite{pki96}\\
(h): Tebaldi, Menech, and Stella 1999 \cite{tebaldi99}\\
(i): Majumdar and Dhar 1992 \cite{majumdar92} 
\label{tab2}
\end{table}

\begin{table}

\caption{Comparisons of critical height probabilities $n^*_i$ for
square (sq), plane triangular (pt), and honeycomb (hc) lattices
obtained by 
RG transformation, numerical simulations, and exact calculation. 
In the numerical simulations with statistics of $10^6$ configurations
on $1000 \times 1000$ lattices are generated to obtain the data.}
\begin{tabular}{ccccccc}      

          &  $n_0^*$  &  $n_1^*$  & $n_2^*$  &  $n_3^*$ & $n_4^*$ & $n_5^*$   \\  \hline \hline
RG (sq)$^{(a)}$     &   0.021   & 0.134     &  0.349    &   0.496  & ----- & -----   \\  \hline
RG (sq)$^{(b)}$      &   0.021   & 0.134     &  0.349    &   0.496  & ----- & -----   \\  \hline
Simulation (sq)$^{(a)}$      &   0.074   & 0.174     &  0.306    &   0.446  & ----- & -----   \\  \hline
Exact (sq)$^{(c)}$   &   0.074  &   0.174  &  0.306  &    0.446 & -----  & ----- \\ \hline \hline
RG (pt)$^{(a)}$      &   0.036   & 0.135     &  0.198    &   0.210  & 0.211 & 0.211   \\  \hline
RG (pt)$^{(d)}$     &   0.036   & 0.135     &  0.198    &   0.210  & 0.211 & 0.211   \\  \hline
Simulation (pt)$^{(a)}$   &   0.058  &   0.094  &  0.139  &    0.188 & 0.240  & 0.281 \\ \hline \hline
RG (hc)$^{(a)}$      &   0.014   & 0.308     &  0.678    &   -----  & ----- & -----   \\  \hline
Simulation (hc)$^{(a)}$    &   0.083  &   0.293  &  0.624  &  ----- & -----  & ----- \\ 
\end{tabular}
(a): this work  based on real height RG approach  \\
(b): Ivashkevich 1996 \cite{ivashkevich96} \\
(c): Priezzhev 1994 \cite{priezzhev94} \\
(d): Papoyan and Povolotsky 1997 \cite{papoyan97} \\
\label{tab3}
\end{table}

\begin{table}
\caption{Fixed point of transferring probability $p_i^*$
 and branching probabilities (BP) of spanning trees $\hat{p_i}$}
\begin{tabular}{ccccccc}              
          &  $p_1^*$  ($\hat{p_1})$  &  $p_2^*$  ($\hat{p_2})$ & $p_3^*$  ($\hat{p_3}$)
&  $p_4^*$  ($\hat{p_4}$)& $p_5^*$  ($\hat{p_5}$) & $p_6^*$  ($\hat{p_6}$)  \\  \hline \hline
RG (sq)$^{(a)}$     &   0.295   & 0.435     &  0.229    &   0.041  & ----- & -----   \\  \hline
RG (sq)$^{(b)}$    &   0.295   & 0.435     &  0.229    &   0.041  & ----- & -----   \\  \hline
BP (sq)$^{(c)}$    &   0.295  &   0.447  &  0.222  &    0.036 & -----  & ----- \\ \hline \hline
RG (pt)$^{(a)}$    & 0.0000179   &  0.00226   &  0.0558  &    0.296 & 0.471  & 0.174 \\ \hline
RG (pt)$^{(d)}$  & 0.0000179   &  0.00226   &  0.0558  &    0.296 & 0.471  & 0.174 \\ \hline
BP (pt)$^{(d)}$      & 0.322     & 0.417     &  0.207    &   0.049  & 0.006 & 0.0002   \\  \hline \hline 
RG (hc)$^{(a)}$      &   0.546   & 0.432     &  0.022    &   -----  & ----- & -----   \\  \hline
BP (hc)$^{(a)}$  &   0.25  &   0.5  &  0.25  &  -----  & -----  & ----- \\ 
\end{tabular}
(a): this work based on real height RG approach  \\
(b): Ivashkevich 1996 \cite{ivashkevich96} \\
(c): Manna, Dhar, and Majumdar 1992 \cite{manna92} \\
(d): Papoyan and Povolotsky 1997 \cite{papoyan97} \\
\label{tab4}

\end{table}

\end{document}